\documentclass[a4paper,11pt]{article}
\pdfoutput=1 

\usepackage{jcappub} 

\usepackage[T1]{fontenc} 

\usepackage[numbers]{natbib}
\usepackage{enumitem}

\usepackage{bm}
\let\vec\bm

\newcommand{\diff}{\ensuremath{\mathrm{d}}}
\newcommand{\e}{\mathrm{e}}

\newcommand{\ac}{a_\mathrm{coll}}
\newcommand{\rc}{r_\mathrm{core}}
\newcommand{\Rc}{r_\mathrm{cusp}}
\newcommand{\Mc}{M_\mathrm{cusp}}
\newcommand{\erf}{\mathrm{erf}}
\newcommand{\fmax}{f_\mathrm{max}}
\newcommand{\rhoc}{\bar\rho_0}

\newcommand{\fsurv}{f_\mathrm{surv}}

\title{Prompt cusps and the dark matter annihilation signal}

\author{M. Sten Delos}
\author{and Simon D. M. White}
\affiliation{Max-Planck-Institut für Astrophysik,\\Karl-Schwarzschild-Str. 1, 85748 Garching, Germany}

\emailAdd{sten@mpa-garching.mpg.de}
\emailAdd{swhite@mpa-garching.mpg.de}

\abstract{
As the first dark matter objects gravitationally condense, a density cusp forms immediately at every initial density maximum. Numerical simulations and theoretical arguments suggest that these prompt cusps can survive until the present day. We show that if dark matter is a thermally produced weakly interacting massive particle, many thousands of prompt cusps with individual masses similar to that of the Earth may be present in every solar mass of dark matter. This radically alters predictions for the amount and spatial distribution of dark matter annihilation radiation. The annihilation rate is boosted by at least an order of magnitude compared to previous predictions, both in the cosmological average and within galaxy-scale halos. Moreover, the signal is predominantly boosted outside of the centers of galactic halos, so alternative targets become significantly more attractive for indirect-detection searches. For example, prompt cusps present new opportunities to test the dark matter interpretation of the Galactic Center $\gamma$-ray excess by searching for similar spectral signatures in the isotropic $\gamma$-ray background and large-scale cosmic structure.
}

\begin{document}
\maketitle
\flushbottom

\section{Introduction}
\label{sec:intro}

The first dark matter systems form by direct monolithic collapse of peaks in the smooth density field of the early Universe. At the moment of collapse, a density cusp forms quasi-instantaneously at the center of each peak \cite{2019PhRvD.100b3523D,2023MNRAS.518.3509D}. These prompt cusps have power-law density profiles 
$\rho\propto r^{-1.5}$, with amplitude and extent tightly linked to those of the linear density peaks from which they form \cite{2019PhRvD.100b3523D,2023MNRAS.518.3509D}.\footnote{Density profiles similar to $\rho\propto r^{-1.5}$ were also noted to arise at the centers of the first or smallest dark matter halos by Refs.~\cite{2010ApJ...723L.195I,2013JCAP...04..009A,2014ApJ...788...27I,2015MNRAS.450.2172P,2016MNRAS.461.3385O,2017MNRAS.471.4687A,2018PhRvD..97d1303D,2018PhRvD..98f3527D,2018MNRAS.473.4339O,2020MNRAS.492.3662I,2021A&A...647A..66C}.} This process is effectively self-similar \cite{2022MNRAS.517L..46W,2023PDU....4101259D} and differs qualitatively from the slow build-up of more massive halos, which is intimately related to their non-power-law density profiles \cite{2013MNRAS.432.1103L} well represented by the ``universal'' Navarro-Frenk-White (NFW) form \cite{1996ApJ...462..563N}. 
As a consequence, the properties of the prompt cusp population are set directly by the cosmological initial conditions and cannot be inferred correctly by extrapolating results from simulations of much more massive systems.
Previous work on the evolution of halo substructure indicates that as high-mass halos grow through accretion and merging, almost all the prompt cusps they accumulate survive 
until late times with at least their inner regions intact \cite{2020MNRAS.491.4591E,2023MNRAS.tmp..828S}. Thus, the present-day abundance and structural properties of prompt cusps are set directly by the cosmological initial conditions \cite{1986ApJ...304...15B}; they are the oldest and densest elements of dark matter structure.

The most widely discussed class of dark matter models involves weakly interacting massive particles (WIMPs) that were thermally produced in the early universe and can annihilate today into detectable radiation \cite{2018RPPh...81f6201R,2018EPJC...78..203A}.
Numerous indirect-detection experiments have searched for annihilation products, typically $\gamma$ rays.
As a two-body process, the annihilation rate within each volume element is proportional to the square of the local dark matter density.
This means that the presence of high-density regions, like prompt cusps, can greatly amplify annihilation.

Previous annihilation searches have largely targeted the nearest galaxy-scale dark matter systems: our own Galactic Center \cite{2017ApJ...840...43A,2022PhRvL.129k1101A}, nearby dark matter-dominated satellite galaxies \cite{2015PhRvL.115w1301A}, and nearby galaxy clusters \cite{2022PhRvD.106j3526B,2023PhRvD.107h3030D}.
This search strategy is motivated by the high dark matter density expected at the centers of these systems.
Intriguingly, an excess of $\gamma$ rays originating from the Galactic Center has been identified \cite{2011PhLB..697..412H} with an emission morphology that closely matches the prediction from high-resolution numerical simulations \cite{2022MNRAS.511L..55G}.
However, the estimate underlying this prediction neglects the contribution of unresolved dark matter substructure.
Cold dark matter models predict the existence of halos on all scales more massive than the prompt cusps \cite{2004MNRAS.353L..23G}, so galactic halos are expected to contain a vast array of subhalos, the tidally stripped remnants of earlier generations of smaller halos \cite{2019Galax...7...81Z}. Recently, both numerical simulations \cite{2020Natur.585...39W,2021MNRAS.501.3558G} and semianalytic models \cite{2019Galax...7...68A} have been used to extrapolate this substructure down by the 10 or more orders of magnitude needed to predict the properties of the smallest subhalos, estimating that they boost the annihilation rate in galactic systems by  factors of a few.
We will see that this conclusion shifts dramatically when we evaluate directly the properties of the prompt cusps, the smallest and densest substructures.

The prompt cusps predicted in WIMP models boost the annihilation rate inside galactic halos by factors of order 100. Moreover, this boost is associated with a dramatic change to the spatial distribution of dark matter annihilation. Prompt cusps dominate the dark matter annihilation rate everywhere except the dense centers of galactic halos, and where they dominate, the annihilation rate scales with the number of prompt cusps and hence with the average dark matter density, rather than with its square. Consequently, the signal expected for indirect-detection searches is much more broadly extended than has been assumed in past works. For example, if the Galactic Center $\gamma$-ray excess arises from dark matter annihilation, then a similar signal should be seen in the isotropic $\gamma$-ray background and nearby large-scale structure.

\section{Prompt cusps}

We begin by discussing the prompt cusps that are predicted to arise in WIMP scenarios.

\subsection{The peak-cusp connection}\label{sec:peakcusp}

\begin{figure}[t]%
\centering
\includegraphics[width=0.7\textwidth]{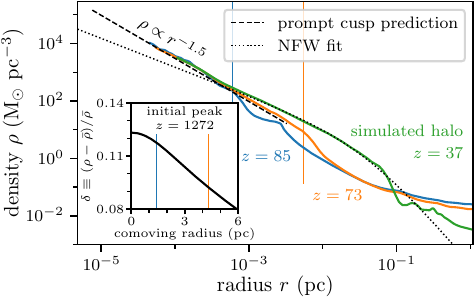}
\caption{
{Prompt cusp formation in a numerical simulation.}
This figure shows the density profile produced by a peak that collapses at $z=87$. A $\rho\propto r^{-1.5}$ density cusp develops immediately. By $z=85$ (blue curve), it already closely matches the amplitude predicted from the properties of the initial linear density field (dashed line extending between the predicted $\rc$ and $\Rc$; this is not a fit), and by $z=73$, it also matches the predicted size.
This prompt cusp persists at the center of the developing halo even at the later redshift $z=37$ (blue curve), when accretion has built up a density profile of NFW form (dotted curve) at larger radii.
The inset plot shows the radial profile of the initial density peak at the much earlier redshift $z=1272$. The fractional density contrast $\delta$ is plotted in this case.
The blue and orange vertical lines in the main plot enclose masses at $z=85$ and $z=73$, respectively, that equal the masses enclosed within the lines of the same color in the inset plot.
The blue lines, in particular, emphasize how small the portion of the initial density peak is that sets the amplitude of the prompt cusp.
} \label{illustrate}
\end{figure}

The properties of each prompt cusp are directly related to those of the linear density peak from which it collapsed \cite{2019PhRvD.100b3523D,2022MNRAS.517L..46W,2023MNRAS.518.3509D}. 
To motivate why this connection works, Fig.~\ref{illustrate} shows the density profile of the object formed by a peak that collapses at redshift $z\simeq 87$.
We assume a WIMP with mass $m_\chi=100$~GeV that kinetically decouples from the Standard Model radiation at a temperature of $T_\mathrm{kd}=30$~MeV, typical parameters for supersymmetric dark matter \cite{2004MNRAS.353L..23G}.
By $z=85$ (blue curve), a prompt $\rho=A r^{-1.5}$ cusp is already present. It formed so quickly after peak collapse that its properties can be sensitive only to the immediate neighborhood of the initial density peak (to the left of the blue line in the inset plot).
The broader environment is not yet part of the collapsed structure, so it can only affect the evolution through tidal forces.
The cusp is 10 times larger in radius by $z=73$ (orange curve) but still maintains the amplitude set during the initial collapse.
At a later redshift $z=37$ (green curve), accretion has built up an NFW profile at larger radii (dotted curve), but the prompt cusp persists unaltered at the center of the system.

This behavior is the basis for the universal peak-cusp connection developed by Refs.~\cite{2023MNRAS.518.3509D,2019PhRvD.100b3523D}.
Given an initial peak in the density contrast field, $\delta\equiv(\rho-\bar\rho)/\bar\rho$, the prompt cusp it forms has density profile $\rho=Ar^{-1.5}$ with
\begin{equation}\label{A}
    A\simeq 24\rhoc \ac^{-1.5} R^{1.5}.
\end{equation}
Here, $\rhoc$ is the mean dark matter density today, $\ac$ is the cosmic expansion factor at the peak's collapse, and $R\equiv |\delta/\nabla^2\delta|^{1/2}$ is the peak's characteristic comoving size. This cusp extends out to a radius
\begin{equation}\label{Rc}
    \Rc \simeq 0.11\ac R.
\end{equation}
Finally, there is an interior radius $\rc$ at which the $\rho=A r^{-1.5}$ cusp gives way to a central core \cite{2023MNRAS.518.3509D}. This feature is enforced by Liouville's theorem, which implies that the maximum phase-space density $\fmax$, set in the early universe, cannot be exceeded anywhere in later nonlinear structures \cite{1979PhRvL..42..407T}. For dark matter that decoupled from the radiation while it was nonrelativistic at scale factor $a_\mathrm{kd}$ and temperature $T_\mathrm{kd}$,
\begin{equation}\label{fmax}
    \fmax = (2\pi)^{-3/2}(m_\chi/T_\mathrm{kd})^{3/2}\rhoc a_\mathrm{kd}^{-3},
\end{equation}
and the phase-space structure of a $\rho=A r^{-1.5}$ cusp implies that $\fmax$ is reached at the radius $\rc$ where
\begin{equation}\label{rhor6}
    A \rc^{4.5} \simeq 3\times 10^{-5} G^{-3}\fmax^{-2}.
\end{equation}
For WIMP dark matter scenarios, $\Rc/\rc\sim 500$ for most prompt cusps, which implies that the $\rho\propto r^{-1.5}$ density profile extends over about that factor in radius. At radii $r\ll \rc$, a prompt cusp should possess uniform density $\rho\simeq A\rc^{-1.5}$.
The dashed line in Fig.~\ref{illustrate} shows the $\rho=Ar^{-1.5}$ density profile predicted from the properties of this particular initial linear density peak; it is plotted between the predicted radii $\rc$ and $\Rc$. It closely matches the density profile of the simulated halo at $z=73$ (orange curve) down to  the simulation's resolution limit.\footnote{Figure~\ref{illustrate} shows the simulated halo C3 from Ref.~\cite{2023MNRAS.518.3509D}, scaled to match our WIMP scenario as described in that article. The density profile is plotted down to 4 times the force-softening length, a radius that encloses 800 particles at $z=85$.}
It should be interpreted as a lower limit on the density near $\Rc$, because later accretion can increase the density at intermediate radii,  as shown by the green curve.

\subsection{The prompt cusp distribution}\label{sec:dist}

\begin{figure}[b]%
\centering
\includegraphics[width=0.7\textwidth]{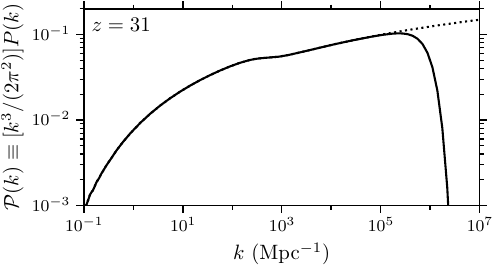}
\caption{{Dimensionless power spectrum $\mathcal{P}(k)$ of dark matter density variations, evaluated in linear theory at $z=31$ in a 100~GeV WIMP scenario with $T_\mathrm{kd}=30$~MeV.} This model has free-streaming wavenumber $k_\mathrm{fs}=1.06\times 10^6$~Mpc$^{-1}$. The dotted line shows the power spectrum if the dark matter is instead arbitrarily cold. The feature between $10^2$ and $10^3$~Mpc$^{-1}$ arises because at larger scales, baryons cluster with the dark matter, while at smaller scales, baryons are a homogeneous background. At linear order, dark matter perturbations $\delta$ grow as $\delta\propto a$ in the former case and at the slightly slower rate $\delta\propto a^{0.9}$ in the latter case.}\label{power}
\end{figure}

Due to this peak-cusp connection, it is possible to evaluate the distribution of prompt cusps directly from the statistics of post-recombination linear density fluctuations at early times. These are specified by their power spectrum $P(k)$. Adopting cosmological parameters as estimated by the Planck mission \cite{2020A&A...641A...6P}, we compute the power spectrum of cold dark matter at $z=31$ at linear order using the \textsc{CLASS} Boltzmann solver \cite{2011JCAP...07..034B}, extrapolating it to unresolved small scales using the small-scale linear growth solution valid in mixed matter-radiation domination given in Ref.~\cite{1996ApJ...471..542H}. 
This yields the dotted curve in Fig.~\ref{power}.
We remark that the feature between $k\simeq 10^2$~Mpc$^{-1}$ and $k\simeq 10^3$~Mpc$^{-1}$ is associated with baryonic physics. At larger scales $k < 10^{2}$~Mpc$^{-1}$, baryons cluster with the dark matter, and perturbations grow at the usual rate $\delta\propto a$ in linear theory, where $a$ is the cosmic expansion factor. At smaller scales $k> 10^{3}$~Mpc$^{-1}$, baryons resist clustering due to their pressure \cite{2006PhRvD..74f3509B}, and dark matter perturbations grow at the rate $\delta\propto a^{g}$ with $g\simeq 0.901$ \cite{1996ApJ...471..542H}. When evaluating the prompt cusp distribution, we will approximate that $\delta\propto a^{0.901}$ at all the (small) scales of interest.

Next, we specialize to a WIMP dark matter model. As an example, we fix the particle mass to be $m_\chi=100$~GeV and the temperature at kinetic decoupling to be $T_\mathrm{kd}=30$~MeV, the same parameters used for Fig.~\ref{illustrate}. The residual thermal motion of the dark matter smooths the density field on a characteristic free-streaming scale $k_\mathrm{fs}^{-1}$. We evaluate $k_\mathrm{fs}$ and scale the power spectrum by $\exp(-k^2/k_\mathrm{fs}^2)$ according to Ref.~\cite{2006PhRvD..74f3509B}. For the WIMP model under consideration, $k_\mathrm{fs}=1.06\times 10^6$~Mpc$^{-1}$, and the resulting linear power spectrum of dark matter density variations is depicted by the solid curve in Fig.~\ref{power}.

\begin{figure}[t]%
\centering
\includegraphics[width=\textwidth]{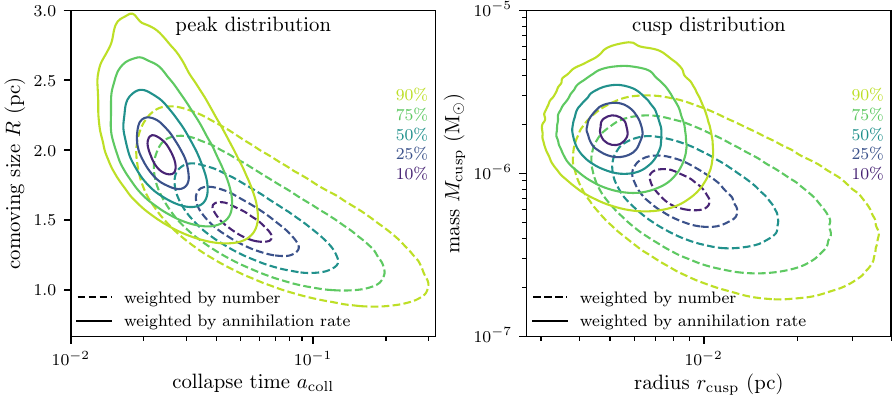}
    \caption{
    {Distribution of peaks (left-hand panel) and the resulting prompt cusps (right-hand panel) in the 100~GeV WIMP scenario with $T_\mathrm{kd}=30$~MeV.}
    We show contours that enclose 10, 25, 50, 75, and 90 percent of the distribution. Dashed contours are weighted by the number of cusps, while solid contours are weighted by the annihilation rate. The cusp sizes $R$ tend to lie within a factor of a few of the free-streaming scale $k_\mathrm{fs}^{-1}=0.94$~pc. Prompt cusps arising from earlier-forming and larger initial peaks tend to dominate the annihilation rate; they end up at the high end of the cusp mass distribution and the low end of the radius distribution.
    }\label{dist2d}
\end{figure}

With the dark matter power spectrum established, we can use the well-understood statistics of the population of initial density peaks \cite{1986ApJ...304...15B} to quantify the statistics of the prompt cusps. This calculation is detailed in Appendix~\ref{sec:stats}.
For a primordial power spectrum with Planck parameters, our 100~GeV WIMP scenario predicts about $10^5$ peaks in the initial density field per solar mass of dark matter.
We estimate a collapse time $\ac$ for each  peak using the ellipsoidal collapse approximation of Ref.~\cite{2001MNRAS.323....1S}.
This approximation predicts that about half of the peaks will actually collapse. The left-hand panel of Fig.~\ref{dist2d} shows how those peaks are distributed in collapse time $\ac$ and size $R\equiv |\delta/\nabla^2\delta|^{1/2}$. For each peak, we use Eqs. (\ref{A}) and~(\ref{Rc}) to evaluate the properties of the resulting prompt cusp. The right-hand panel of Fig.~\ref{dist2d} shows the distribution of cusps in mass $\Mc$ and radius $\Rc$, where $\Mc=(8\pi/3)A\Rc^{1.5}$ is the mass enclosed within $\Rc$.
Weighted by contribution to the annihilation signal, the average prompt cusp has $A\simeq 6\times 10^{-4}$~M$_\odot$pc$^{-1.5}$, $\rc\simeq 1.2\times 10^{-5}$~pc, $\Rc\simeq 6\times 10^{-3}$~pc, and a mass of $2.5\times 10^{-6}$~M$_\odot$.

\subsection{Annihilation in prompt cusps}

\begin{figure}[t]
\centering
\includegraphics[width=0.7\textwidth]{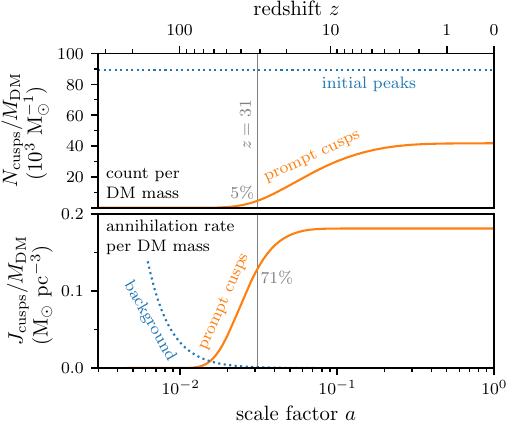}
\caption{
{The prompt cusp population as a function of time for a 100~GeV WIMP that decouples at 30~MeV, assuming that each collapsed peak can be associated with a cusp.}
Normalized to the mass $M_\mathrm{DM}$ of dark matter (DM), the upper panel shows the cusp count $N_\mathrm{cusps}$, while the lower panel shows the total volume-integrated squared density in cusps, $J_\mathrm{cusps}$, which is proportional to the total annihilation rate.
The dotted line in the upper panel shows the number of initial peaks. 
By redshift $z=31$ (gray vertical line), 5 percent of the initial peaks have collapsed to form prompt cusps, but those cusps already contribute 71 percent of the annihilation luminosity of the full distribution.
The lower panel also shows the annihilation rate predicted for dark matter at the mean cosmological density (dotted curve).
}\label{history}
\end{figure}

Assuming that every collapsed peak is associated with a prompt cusp at later times, the upper panel of Fig.~\ref{history} shows how the predicted cusp count grows over time. At late times, $\sim 40000$ cusps are predicted per solar mass of dark matter.
While the low masses and extended nature of these prompt cusps make them undetectable by gravitational means, they are dense enough that they greatly boost the total amount of dark matter annihilation.
The annihilation rate inside a volume is proportional to
\begin{equation}\label{J}
    J\equiv\int\rho^2 \diff V
\end{equation}
integrated over the volume.
We assume that prompt cusps have the density profile specified by Ref.~\cite{2023MNRAS.523.1067S}, which is a $\rho=Ar^{-1.5}$ density cusp modified in phase space so as not to exceed the density $\fmax$.
The annihilation $J$ factor inside a prompt cusp is
\begin{equation}\label{Jcusp}
    J_\mathrm{cusp} = 4\pi A^2\left[0.531 + \log(\Rc/\rc)\right]
\end{equation}
for this profile.
We integrate $J_\mathrm{cusp}$ over the cusp distribution to obtain the aggregate annihilation signal. The lower panel of Fig.~\ref{history} shows how the aggregate $J_\mathrm{cusps}$ grows over time.
In particular, we find that 5 percent of the initial peaks form their prompt cusps by redshift $z=31$, but those cusps already account for 71 percent of the annihilation luminosity predicted from the full array of peaks.
0.8 percent of the dark matter resides in these early-forming cusps.

For a WIMP with mass $m_\chi$ that decouples from the radiation at temperature $T_\mathrm{kd}$, we find that the late-time value of the volume-averaged annihilation $J$ factor per unit dark matter mass $M_\mathrm{DM}$ can be approximated as
\begin{equation}\label{JperM_particle}
    \frac{J_\mathrm{cusps}}{M_\mathrm{DM}} = 0.08\fsurv\left[\log\left(\frac{m_\chi}{\mathrm{GeV}}\frac{T_\mathrm{kd}}{\mathrm{GeV}}\right)+36\right]^5 \rhoc,
\end{equation}
where $\rhoc$ is the mean dark matter density today; this expression is accurate to within about 10 percent for WIMPs that decouple above a keV. We have inserted a factor $\fsurv\leq 1$ to account for the fact that not all initial peaks produce prompt cusps that survive up to the present day; as we will discuss below, $\fsurv\simeq 0.5$ (but Fig.~\ref{history} approximates $\fsurv=1$). A less precise but more evocative expression is
\begin{equation}\label{JperM}
    J_\mathrm{cusps}/M_\mathrm{DM}\simeq 160 \fsurv\rhoc (1+z_5)^3,
\end{equation}
where $z_5$ is the redshift by which 5 percent of the initial peaks have collapsed. Thus the prompt cusp population produces an annihilation luminosity per unit mass which is $160\fsurv$ times the prediction for uniformly distributed dark matter at $z_5$. This redshift ranges from 25 to 50 for WIMPs that decouple at MeV or higher temperatures, so prompt cusps dominate the annihilation rate from all present-day regions where they survive and where the dark matter density is less than a few million times the current cosmic mean.

\subsection{Survival of prompt cusps}

While these estimates have so far assumed that every peak forms a prompt cusp and that all prompt cusps survive, the overall annihilation rate is not greatly altered by accounting 
properly for the hierarchical build-up of structure. When halos of very different mass merge, the smaller halo persists as a tidally stripped subhalo orbiting within the larger one, so the prompt cusps of both halos survive. Both theoretical descriptions \cite{2021arXiv211101148A,2022MNRAS.517.1398B,2023MNRAS.tmp..828S,2022MNRAS.516..106D} and numerical simulations \cite{2020MNRAS.491.4591E} suggest that although tidal forces gradually strip mass from the outer regions of subhalos, both central and satellite prompt cusps survive this process almost unaltered.
We will show in Section~\ref{sec:boost} that tidal stripping has only a minor effect on the annihilation rate in prompt cusps.
Prompt cusps that are satellites of galactic systems can be destroyed by stellar encounters, but this predominantly affects cusps orbiting within the innermost star-dominated regions of galaxies and has little effect on the bulk of the population, which lies in their massive halos \cite{2023MNRAS.523.1067S}.

There are nevertheless two main effects that can reduce the number of prompt cusps below that predicted from the distribution of initial peaks. 
Merging of halos of comparable mass can cause their central cusps to merge also, and additionally, some cusps fail to form because their peaks are incorporated in other systems before they collapse.
We use the halo catalogues and merger trees associated with the simulations of Ref.~\cite{2023MNRAS.518.3509D} to estimate that these effects reduce both the number of prompt cusps and their annihilation signal by 40 to 60 percent; that is, $\fsurv\simeq 0.5\pm 0.1$ in Eqs. (\ref{JperM_particle}) and~(\ref{JperM}). This estimate is detailed in Appendix~\ref{sec:simsurv} and rests on the notion that although the complete disruption of low-mass subhalos is common in numerical simulations, it is mostly a numerical artifact \cite{2018MNRAS.474.3043V}; the central cusps of subhalos orbiting within a smooth host potential should always survive \cite{2020MNRAS.491.4591E}. Therefore, a subhalo's prompt cusp can be assumed to survive unless the subhalo is disrupted very close to the center of its host.
While acknowledging that further study is needed to predict precisely the fraction of the prompt cusp annihilation signal that survives hierarchical clustering, we will adopt $\fsurv= 0.5\pm 0.1$ for the remainder of this article.

\section{Boost to the annihilation rate}\label{sec:boost}

Prompt cusps contribute the vast majority of the annihilation signal inside galaxy-scale halos. To quantify this, we evaluate the annihilation boost factor $B$, defined as the ratio of the annihilation rate in prompt cusps to that associated with the halo's smooth dark matter distribution alone. 
If the prompt cusp distribution inside a host halo is the same as the distribution in the field, then this ratio is just $B=(J_\mathrm{cusps}/M_\mathrm{DM})/(J_\mathrm{halo}/M_\mathrm{halo})$, where $M_\mathrm{halo}$ is the host halo's mass and $J_\mathrm{halo}$ is its annihilation $J$ factor.
In practice, prompt cusps inside larger halos are gradually stripped by tidal forces, which reduces their annihilation signal. We use a recent model of tidal stripping \cite{2023MNRAS.tmp..828S} to account for this consideration, as described in Appendix~\ref{sec:tidal}.

\begin{figure}[b]%
\centering
\includegraphics[width=0.7\textwidth]{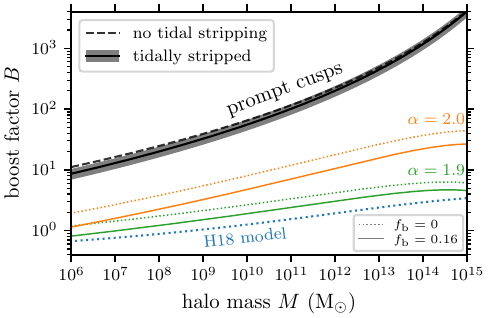}
\caption{{Annihilation boost factor for host halos due to prompt cusps.} We consider again the 100~GeV WIMP model with $T_\mathrm{kd}=30$~MeV. The dashed curve assumes that the prompt cusp signal is scaled by $\fsurv=0.5$ to account for cusp loss during early hierarchical clustering but is unaffected by tidal forces at late times within the final host halo. The solid curve additionally accounts for tidal stripping of the prompt cusps. For the latter case, the band corresponds to $\fsurv=0.5\pm 0.1$.
For comparison, we also include predictions made by extrapolating subhalo abundance models calibrated at much higher subhalo mass. The orange and green curves assume the subhalo mass function to be a power law, $\diff N_\mathrm{sub}/\diff M_\mathrm{sub}\propto M_\mathrm{sub}^{-\alpha}$ \cite{2014MNRAS.442.2271S,2015JCAP...09..008F}. The dotted curves follow previous work in neglecting the early suppression of growth due to baryons (see Fig.~\ref{power}), while the solid curves correctly include this effect, following the assumptions we made when evaluating the prompt cusp distribution.
We also show in blue the prediction of a more recent semianalytic subhalo model \cite{2018PhRvD..97l3002H,2019Galax...7...68A} (H18); this also neglected the early suppression of growth by baryons.}\label{boost}
\end{figure}

We consider host halos with NFW density profiles,
\begin{equation}\label{NFW}
    \rho(r)=\frac{\rho_s}{r/r_s (1+r/r_s)^{2}},
\end{equation}
where the scale radius $r_s$ and scale density $\rho_s$ are related according to the concentration-mass relationship in Ref.~\cite{2014MNRAS.442.2271S}. 
The solid black curve in Fig.~\ref{boost} shows the boost factor $B$ from prompt cusps at $z=0$ as a function of the host halo mass $M$ for the WIMP model with $m_\chi=100$~GeV and $T_\mathrm{kd}=30$~MeV. The shaded gray region represents the uncertainty range in $\fsurv$. Evidently, prompt cusps boost the annihilation rate by a factor of around 30 for halos typical of dwarf galaxies to $\mathcal{O}(10^3)$ for cluster-scale halos. The dashed curve shows the boost factors if tidal stripping is neglected; its impact is marginal.

\subsection{Comparison with conventional halo models}

The annihilation boost factors arising from prompt cusps are much larger than the factors inferred from subhalo models that neglect the prompt cusps. We show several examples in Fig.~\ref{boost}. First, we consider the prescription employed by the Fermi collaboration in Ref.~\cite{2015JCAP...09..008F}. In this prescription, the differential count $N_\mathrm{sub}$ of subhalos of mass $M_\mathrm{sub}$ inside a host halo of mass $M$ is assumed to be a power law, $\diff N_\mathrm{sub}/\diff M_\mathrm{sub}=(A_\mathrm{sub}/M)(M_\mathrm{sub}/M)^{-\alpha}$. Motivated by cosmological simulations \cite{2007ApJ...657..262D,2008MNRAS.391.1685S}, as suggested by  Ref.~\cite{2014MNRAS.442.2271S}, the choices $(A_\mathrm{sub},\alpha)=(0.03,1.9)$ and $(A_\mathrm{sub},\alpha)=(0.012,2)$ are considered. The minimum subhalo mass is assumed to be $M_\mathrm{min}=10^{-6}$~M$_\odot$, approximately that expected for the 100~GeV WIMP model that we have been considering. Subhalos and host halos are assumed to have NFW density profiles according to the concentration-mass relationship in Ref.~\cite{2012MNRAS.423.3018P}. If a halo of mass $M$ has $J$ factor $J(M)$, the annihilation boost factor $B(M)$ is
\begin{equation}\label{subhaloboost}
B(M) = \frac{1}{J(M)}\int_{M_\mathrm{min}}^M
\diff M_\mathrm{sub} \frac{\diff N_\mathrm{sub}}{\diff M_\mathrm{sub}}
[1+B(M_\mathrm{sub})] J(M_\mathrm{sub}),
\end{equation}
which we evaluate iteratively starting with $B=0$. Convergence is achieved after 4 iterations, although we proceed to 8.

The orange curves in Fig.~\ref{boost} show the subhalo model predictions for $\alpha=2$, while the green curves show the predictions for $\alpha=1.9$. The dotted curves neglect the early suppression of growth by baryons (depicted in Fig.~\ref{power}), which is the approximation made by Refs.~\cite{2014MNRAS.442.2271S,2015JCAP...09..008F} and most previous studies of the annihilation boost; to produce these curves, we evaluated halo concentrations using the power spectrum provided by Ref.~\cite{1998ApJ...496..605E}, which does not account for baryonic growth suppression at small scales. However, our evaluation of the signal from prompt cusps accounted correctly for this effect, as we discussed in Section~\ref{sec:dist}. To provide a more accurate comparison, the solid orange and green curves in Fig.~\ref{boost} therefore account for a nonzero baryon fraction $f_\mathrm{b}=0.16$. For these curves, we evaluate halo concentrations using a linear dark matter power spectrum evaluated at $z=0$ using the CLASS code (similar to that shown in Fig.~\ref{power}), instead of using the power spectrum in Ref.~\cite{1998ApJ...496..605E}. However, we warn that the halo concentration model of Ref.~\cite{2012MNRAS.423.3018P} was not calibrated using simulations including this effect.

Evidently, the annihilation boost from prompt cusps is at least an order of magnitude higher than that predicted by these subhalo models for corresponding assumptions. Moreover, the subhalo models are intrinsically uncertain because they extrapolate numerical results down to mass scales more than 10 orders of magnitude below the resolution limits of the simulations they are based on.
Non-halo-based approaches to substructure modeling have also been considered \cite{2010PhRvD..81d3532K,2012MNRAS.421L..87S,2016MNRAS.457..986Z}, but they tend to rely on similar extrapolations.
As an alternative, we compare to the predictions of the recent semianalytic subhalo model of Refs.~\cite{2018PhRvD..97l3002H,2019Galax...7...68A} (H18), which builds up the subhalo population over cosmic time by modeling in detail the accretion and evolution of these objects. The blue dotted curve in Fig.~\ref{boost} shows the boost factors $B$ predicted from this model.\footnote{The boost factors here are slightly larger than those presented in Ref.~\cite{2019Galax...7...68A} because we set subhalo accretion to begin at $z=50$ instead of $z=7$.} This model also neglects early suppression of small-scale growth due to baryons, but it nevertheless predicts even lower annihilation boost factors.

\begin{figure}[t]%
\centering
\includegraphics[width=0.7\textwidth]{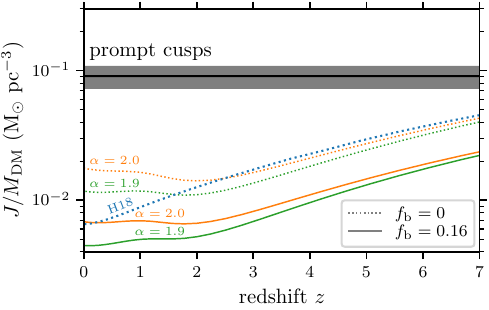}
\caption{{Predictions for the cosmologically averaged annihilation rate from prompt cusps (black line) and halo models (colors) for the same 100~GeV WIMP model as in Fig.~\ref{boost}.}
The shaded band corresponds to $\fsurv=0.5\pm 0.1$.
As in Fig.~\ref{boost}, the orange and green curves assume that the subhalo mass function is a power law \cite{2014MNRAS.442.2271S} while the blue curve adopts a semianalytic subhalo model \cite{2018PhRvD..97l3002H,2019Galax...7...68A} (H18).
The dotted curves neglect the early suppression of growth due to baryons, while the solid curves account for this effect; the former is the choice made by previous studies, while the latter is more correct and matches the assumption made in evaluating the prompt cusp prediction.}\label{global}
\end{figure}

Figure~\ref{global} compares the cosmologically averaged value of $J/M_\mathrm{DM}$ for prompt cusps and the same halo models considered above. For the halo models, this is obtained by integrating the subhalo-boosted $J$ factors of host halos over the field halo mass function given by Ref.~\cite{2012MNRAS.423.3018P}; this is the same mass function employed in the Fermi collaboration's analysis in Ref.~\cite{2015JCAP...09..008F}. For the prompt cusps, we take $J_\mathrm{cusps}/M_\mathrm{DM}$ in Fig.~\ref{history} and scale it by $\fsurv$.
Evidently, the global annihilation rate from prompt cusps exceeds the zero-baryon halo model predictions today by a factor of at least 5 and the predictions of the halo models with baryons accounted for by over an order of magnitude. Baryons have a larger impact here than in Fig.~\ref{boost} because in addition to the change to the dark matter power spectrum, the overall annihilation rate is scaled by $(1-f_\mathrm{b})^2$. The difference between halo model and prompt cusp predictions is smaller in the past, which could be attributed to tidal stripping of the subhalos inside host halos (although this is only explicitly included in the H18 model); tidal stripping has a much smaller impact on the highly compact prompt cusps than on the subhalos more broadly. However, it should also be noted that the halo models are generally calibrated to be accurate at $z=0$ rather than at high redshift.
The general conclusions that we can draw from these comparisons are that
\begin{enumerate}[label={(\arabic*)}]
    \item prompt cusps greatly boost annihilation compared to previous predictions, and
    \item the annihilation rate today is dominated by the prompt cusps; inclusion of the broader structure of each halo would yield only a small correction.
\end{enumerate}

\subsection{Comparison with modified halo models}

\begin{figure}[b]%
\centering
\includegraphics[width=0.7\textwidth]{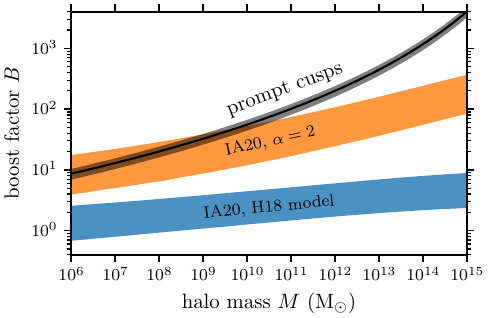}
\caption{
{Comparison between the annihilation boost factors predicted by the prompt cusp paradigm (black) and the predictions of Ref.~\cite{2020MNRAS.492.3662I} (IA20, colored bands), which modeled the smallest halos as possessing NFW profiles with steeper inner cusps.}
We consider the same 100~GeV WIMP model as in Fig.~\ref{boost}.
The orange band uses the same power-law subhalo mass function as the orange curves in Fig.~\ref{boost}, while the blue band uses the semianalytic subhalo model of Refs.~\cite{2018PhRvD..97l3002H,2019Galax...7...68A} (H18), like the blue curves in Fig.~\ref{boost}. The IA20 predictions all neglect the early suppression of small-scale growth caused by  baryons remaining smooth while dark matter clusters; the predicted boost factors would be lower if this effect were accounted for.}\label{boost-compare}
\end{figure}

Going beyond standard halo-model estimates of the annihilation boost, Refs.~\cite{2010ApJ...723L.195I,2014ApJ...788...27I,2020MNRAS.492.3662I} noted that the smallest halos may have steeper inner density profiles than predicted by the NFW paradigm and considered the potential impact of such profiles on boost factors.
Reference~\cite{2020MNRAS.492.3662I} (IA20) carried out boost factor computations by adopting generalized NFW profiles
\begin{equation}
    \rho(r)=\frac{\rho_s}{(r/r_s)^\alpha (1+r/r_s)^{3-\alpha}}
\end{equation}
with inner slopes $\alpha\geq 1$ that depend on halo mass (where $\alpha=1$ corresponds to NFW; compare Eq.~\ref{NFW}). The slopes $\alpha$ and halo concentrations were then tuned to match the average inner density profiles of simulated very-low-mass halos.
Figure~\ref{boost-compare} compares our prediction for the boost factors due to prompt cusps to two results presented in IA20, one adopting the $\alpha=2$ power-law subhalo mass function (orange band) and the other employing the semianalytic subhalo model of Refs.~\cite{2018PhRvD..97l3002H,2019Galax...7...68A} (blue band).
The thickness of the bands accommodates a range of assumptions regarding the inner density profiles of low-mass subhalos.
The annihilation boost from prompt cusps exceeds that for the $\alpha=2$ model by a factor of a few for galaxy-scale halos, and it exceeds that for the semianalytic model by more than an order of magnitude. Note that both models neglect early growth suppression due to baryons; their predictions would be even lower if this effect were properly accounted for (as in the prompt cusp prediction).

While prompt cusps are indeed responsible for the steep density profiles of the simulated ``microhalos'' considered by Refs.~\cite{2010ApJ...723L.195I,2014ApJ...788...27I,2020MNRAS.492.3662I} and others, the prompt cusp population that we consider here is characterized quite differently.
Previous work treated the $\rho\propto r^{-1.5}$ cusp as a simple modification of the NFW density law and analyzed low-mass populations by extrapolating results from simulations of much more massive halos and subhalos. As we have emphasized, however, prompt cusps and NFW halos have different physical origins; the size and amplitude of an early halo's central prompt cusp are not tied to the parameters of its NFW profile but rather to those of the corresponding initial density peak. Previous work on the evolution of first-generation halos in a cosmological context concluded that steep central cusps soften to the NFW form as halos grow \cite{2014ApJ...788...27I,2016MNRAS.461.3385O,2017MNRAS.471.4687A,2019PhRvD.100b3523D,2020MNRAS.492.3662I}, a behavior typically attributed to violent relaxation in halo mergers \cite{2016MNRAS.461.3385O,2017MNRAS.471.4687A,2019PhRvD.100b3523D}. However, this conclusion is likely
a result of numerical resolution limitations, together with plotting and averaging choices, since the significantly higher resolution simulations of Ref.~\cite{2023MNRAS.518.3509D} found initial prompt cusps to persist as NFW halos build up around them (see Fig.~\ref{illustrate}), with even major halo mergers affecting inner density profiles to only a minor degree. As a halo approaches galactic scale, its central cusp is, in any case, vastly outnumbered by all the other prompt cusps it has accreted.
Our prompt cusp analysis here accounts for these considerations by calculating the relevant properties of the population directly from the linear initial conditions.

\section{Implications for dark matter annihilation searches}

The large annihilation boost factors discussed in Section~\ref{sec:boost} are related to a major change in the spatial distribution predicted for the dark matter annihilation signal.
Prompt cusps do not boost the signal from the densest regions, such as the Galactic Center, but they greatly enhance the signal from sparser regions such the outer Galactic halo, galaxy clusters, and the extragalactic field.
We demonstrate this point by considering the annihilation signals that should arise from prompt cusps if the observed Galactic Center $\gamma$-ray excess is due to dark matter annihilation.

\begin{figure}[t]%
\centering
\includegraphics[width=0.7\textwidth]{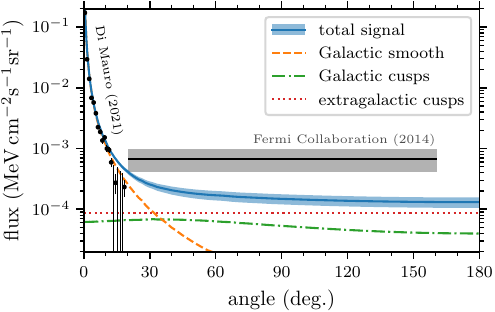}
\caption{
{Energy flux in 1-10 GeV $\gamma$-rays from dark matter annihilation, as a function of angle from the Galactic Center, if the observed $\gamma$-ray excess is due to dark matter annihilation.}
We include the contributions from the smooth Galactic halo (dashed curve) and from the Galactic (dot-dashed curve) and extragalactic (dotted curve) prompt cusp populations.
The blue solid curve shows the predicted total dark matter signal, and we also mark the $\fsurv=0.5\pm 0.1$ uncertainty band.
While the predicted signal matches the measured Galactic Center excess (black points) at angles below 10~degrees, there is slight tension at larger angles.
Above 20~degrees, we compare to the measured isotropic diffuse $\gamma$-ray background \cite{2015ApJ...799...86A} (black line with gray uncertainty band). Dark matter annihilation must evidently account for a significant fraction of this background in the 1-10~GeV energy range if the Galactic Center excess is due to dark matter annihilation.
} \label{galactic}
\end{figure}

The black points in Fig.~\ref{galactic} depict the observed Galactic Center excess (GCE), as reported by Ref.~\cite{2021PhRvD.103f3029D}. Specifically, we show the energy flux in 1-10 GeV $\gamma$ rays.
Under the assumption that this observation can be attributed to dark matter annihilation, the blue curve shows the predicted flux in the same energy range, including both prompt cusps and the smooth halo, as a function of angular distance from the Galactic Center.
For the prompt cusp contribution, we adopt the WIMP scenario with $m_\chi=100$~GeV (consistent with the Galactic Center emission spectrum \cite{2015JCAP...05..011A}) and $T_\mathrm{kd}=30$~MeV, and we use the recent \textsc{cusp-encounters} model \cite{2023MNRAS.523.1067S} to account for the disruption of prompt cusps in the inner Galactic halo by tidal forces and by encounters with individual stars.
Within the central 10 degrees, the smooth Galactic halo (dashed curve) dominates the signal, producing a morphology that closely matches the measurement of Ref.~\cite{2021PhRvD.103f3029D}.
Prompt cusps contribute significantly to the annihilation signal at larger angles, however, placing the predicted signal (solid blue curve) in weak tension with the measurement of the GCE between 10 and 20 degrees.
We also show (gray band) the isotropic diffuse $\gamma$-ray background measured by the Fermi Collaboration \cite{2015ApJ...799...86A} at Galactic latitudes $|b|>20$~degrees, integrated over the same 1-10 GeV energy range.
If the GCE is due to dark matter annihilation, then annihilation must account for a significant fraction of the entire 1 to 10 GeV background. This is in tension with claims that almost all of this background is produced by unresolved discrete extragalactic sources such as star-forming galaxies and active galactic nuclei \cite{2019JCAP...03..019B}.

\begin{figure}[t]%
\centering
\includegraphics[width=0.7\textwidth]{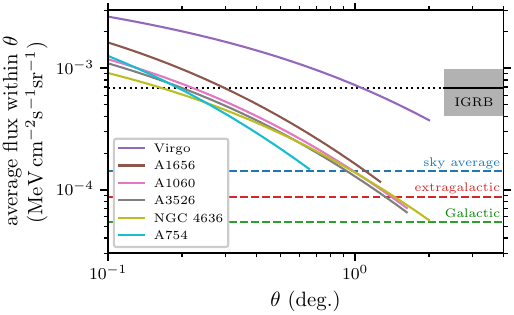}
\caption{
{Energy flux in 1-10 GeV $\gamma$-rays from dark matter annihilation, averaged within the angle $\theta$ of the centers of several low-redshift galaxy clusters, if the observed Galactic Center $\gamma$-ray excess is due to dark matter annihilation.}
We adopt $\fsurv=0.5$. For comparison, the blue dashed line shows the sky-averaged signal from prompt cusps, while the red and green dashed lines show respectively the extragalactic and Galactic contributions thereto.
The dotted line shows the isotropic $\gamma$-ray background (IGRB) measured by the Fermi Collaboration \cite{2015ApJ...799...86A}; the uncertainty band is displayed on the right.
Within a few tenths of a degree of the centers of these clusters, the predicted flux from prompt cusps exceeds the sky average by a factor of a few and even exceeds the total measured IGRB.
} \label{clusters}
\end{figure}

Another possibility is to search for an annihilation signal originating from resolved galaxy clusters (e.g. \cite{2022PhRvD.106j3526B,2023PhRvD.107h3030D}). Assuming again that the GCE is due to dark matter annihilation, Fig.~\ref{clusters} shows the energy flux from $\gamma$ rays in the 1-10 GeV energy range averaged within the angle $\theta$ of the centers of several of the galaxy clusters that were analyzed by Ref.~\cite{2023PhRvD.107h3030D}. For comparison, we also show the sky average of the background flux considered in Fig.~\ref{galactic}.
For these clusters, the flux from prompt cusps averaged within the central few tenths of a degree (approximately the angular resolution of the Fermi LAT at the relevant energies) evidently exceeds the background contribution from prompt cusps in our own and other halos by a factor of a few. It can even exceed the total measured isotropic diffuse background (dotted line).

\section{Conclusion}

Prompt cusps enable new tests of the WIMP hypothesis.
These structures arise generically in collisionless dark matter models in which the initial density field is smooth below some scale. They form at the moment of collapse of initial density maxima. Although they are too small to produce significant gravitational signatures (unless the dark matter is warm \cite{2023MNRAS.522L..78D}), the great abundance and high internal density of prompt cusps make them dominate the dark matter annihilation signal at late times. We find that the inclusion of prompt cusps increases the annihilation rate relative to previous halo/subhalo-based estimates by at least an order of magnitude.

Because prompt cusps dominate the annihilation rate in all but the very densest regions, the annihilation luminosity from most regions is proportional to the number of prompt cusps they contain, hence to the mean dark matter density within them, rather than to the square of this density. The spatial templates typically used in searches for annihilation signals focus on the densest inner regions, and are hence suboptimal; a more sensitive strategy would use broader templates enclosing all regions of high surface mass density (and hence high annihilation surface brightness). 
For example, if the Galactic Center $\gamma$-ray excess is produced by annihilating dark matter, then a significant fraction of the isotropic $\gamma$-ray background should also come from prompt cusps in our own and other galaxy halos.
A new evaluation of the constraints on dark matter particle properties implied by the apparent absence of such a signal is the subject of Ref.~\cite{2023arXiv230713023D}.
Observed large-scale structure presents another promising target; the surface brightness from prompt cusps within the central few tenths of a degree of nearby galaxy clusters is predicted to exceed the total isotropic annihilation signal by a factor of a few.
Regardless of the interpretation of the GCE, the inclusion of prompt cusps will tighten limits on annihilation cross sections from $\gamma$-ray observations of a broad variety of potential sources.

Like any substructure-boosted annihilation rate, the annihilation rate in prompt cusps is sensitive to the cosmological initial conditions at smaller scales than have been probed by observations.
For example, it is logarithmically sensitive to the temperature at which dark matter and radiation decoupled (see Eq.~\ref{JperM_particle}), since that determines the free-streaming scale of the dark matter. However, different decoupling temperatures can shift our conclusions only moderately; dark matter candidates that decouple at sub-MeV temperatures are susceptible to laboratory detection \cite{2013PhRvD..88a5027C}.
More drastic changes to the annihilation rate are possible if the primordial power spectrum at small scales differs substantially from the inflation-motivated, power-law form usually used to extrapolate from cosmic microwave background measurements \cite{2020A&A...641A...6P} or if the thermal history of the early Universe differs substantially from the simplest assumptions (e.g.~\cite{2021OJAp....4E...1A}).
Unlike previous approaches based on subhalo modeling, the prompt cusp paradigm enables straightforward evaluation of the annihilation rate for alternative initial conditions of this kind (e.g.~\cite{2019PhRvD.100l3546D,2023PhRvD.108b3528D}).





\section*{Code availability}

Code to evaluate prompt cusp distributions is available from MSD on request. However, it is adapted from code already available at \url{https://github.com/delos/microhalo-models}.

\appendix

\section{Statistics of peaks and prompt cusps}\label{sec:stats}

According to the peak-cusp connection detailed in Section~\ref{sec:peakcusp}, the properties of a prompt cusp are determined by the collapse scale factor $\ac$ and characteristic size  $R\equiv |\delta/\nabla^2\delta|^{1/2}$ of the density peak from which it formed.
If a peak has amplitude $\delta(a)$ at some scale factor $a$, then the scale factor at which it collapses is 
\begin{equation}\label{acoll}
\ac=\left[f_\mathrm{ec}(e,p)\delta_\mathrm{c}/\delta(a)\right]^{1/g}a,
\end{equation}
where $\delta_\mathrm{c}=1.686$ is the critical density contrast for collapse of a spherical perturbation,\footnote{Reference~\cite{2019PhRvD.100l3546D} found that the spherical collapse threshold is still $\delta_\mathrm{c}=1.686$ (in the dark matter) when there is a homogeneous baryon background.} $g\simeq 0.901$ is the growth index at small scales, and $f_\mathrm{ec}(e,p)$ is the ellipsoidal collapse correction, approximated as \cite{2001MNRAS.323....1S}
\begin{equation}\label{ec}
f_\mathrm{ec}(e,p) = 1+0.47\left[5(e^2-p| p|)f_\mathrm{ec}^2(e,p)\right]^{0.615},
\end{equation}
which depends on the ellipticity $e$ and prolateness $p$ of the tidal field at the peak. Note that $f_\mathrm{ec}$ grows to more than 5 as $e^2-p| p|$ approaches 0.26, and Eq.~(\ref{ec}) has no solution when $e^2-p| p| > 0.26$, which is why the count of prompt cusps in Fig.~\ref{history} does not approach the number of peaks. We assume that these peaks do not collapse, but peaks of such high ellipticity have heights $\delta$ that are so low that their prompt cusps would anyway contribute negligible annihilation luminosity.

Given the dark matter power spectrum, we can straightforwardly employ the statistics of peaks to quantify how $\delta$, $\nabla^2\delta$, $e$ and $p$ are distributed \cite{2019PhRvD.100b3523D}. For convenience, we reproduce the relevant equations, which are expressed in terms of $\sigma_j^2\equiv\int\frac{\diff k}{k}\mathcal{P}(k) k^{2j}$, $R_* \equiv \sqrt 3 \sigma_1/\sigma_2$, and $\gamma\equiv \sigma_1^2/(\sigma_0\sigma_2)$. Here $\mathcal{P}(k)\equiv [k^3/(2\pi^2)]P(k)$ is the dimensionless dark matter power spectrum (e.g. Fig.~\ref{power}). The differential number density of peaks, in terms of parameters $\nu\equiv \delta/\sigma_0>0$ and $x\equiv -\nabla^2\delta/\sigma_2>0$, is \cite{1986ApJ...304...15B}
\begin{equation}\label{nux}
\frac{\mathrm{d}^2n_\mathrm{peaks}}{\mathrm{d}\nu\mathrm{d}x}
=
\frac{\mathrm e^{-\nu^2/2}}{(2\pi)^2R_*^3}f(x)\frac{\exp\!\left[-\frac{1}{2}(x-\gamma\nu)^2/(1-\gamma^2)\right]}{[2\pi(1-\gamma^2)]^{1/2}},
\end{equation}
where
\begin{equation}
f(x)
\equiv
\frac{x^3\!-\!3x}{2}\left[\erf\!\left(\sqrt{\frac{5}{2}}x\right)\!+\!\erf\!\left(\sqrt{\frac{5}{8}}x\right)\right]
\!+\!
\sqrt{\frac{2}{5\pi}}\left[\left(\frac{31}{4}x^2\!+\!\frac{8}{5}\right)\mathrm e^{-\frac{5}{8}x^2}\!+\!\left(\frac{x^2}{2}\!-\!\frac{8}{5}\right)\mathrm e^{-\frac{5}{2}x^2}\right].
\end{equation}
Given peak height $\nu$, the distribution of $e$ and $p$ may be approximated as \cite{2001MNRAS.323....1S}
\begin{equation}\label{ep}
f(e,p\mid\nu) = \frac{1125}{\sqrt{10\pi}}e(e^2-p^2)\nu^5\exp\left[-\frac{5}{2}\nu^2(3e^2+p^2)\right];
\end{equation}
this expression is exact for randomly selected points and nearly exact for density peaks.

\begin{figure}[t]%
\centering
\includegraphics[width=0.7\textwidth]{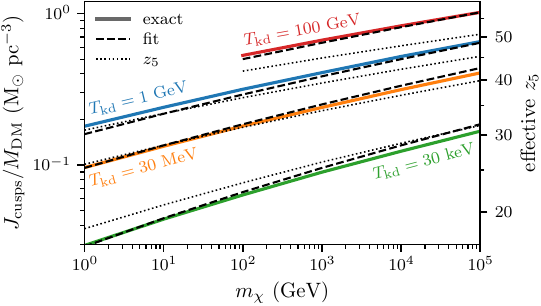}
\caption{
{Influence of the dark matter particle mass $m_\chi$ and the kinetic decoupling temperature $T_\mathrm{kd}$ on the formation epoch of prompt cusps and  on their contribution to dark matter annihilation.}
As a function of $m_\chi$, solid curves show both the factor  $J_\mathrm{cusps}/M_\mathrm{DM}$ (assuming $\fsurv=1$) obtained by summing Eq.~(\ref{J}) over the full cusp population (left axis) and the formation redshift $z_5$ from this factor using Eq.~(\ref{JperM}) (right axis). Dashed curves show the approximation of Eq.~(\ref{JperM_particle}), while dotted curves show the true cusp formation redshift $z_5$, defined as the redshift by which 5 percent (by number) of the initial peaks have collapsed. Comparing the solid curves to the dashed and to the dotted curves thus shows the accuracy of the approximations (\ref{JperM_particle}) and (\ref{JperM}) respectively. There are substantial errors only for the dotted curve at the highest value of $T_\mathrm{kd}$.
}\label{z10}
\end{figure}

We consider a number of WIMP scenarios with different $m_\chi$ and $T_\mathrm{kd}$.
For each scenario, we generate a sample of $10^7$ peaks, using rejection methods to sample $\delta$ and $\nabla^2\delta$ and inverse transforms to sample $e$ and $p$ (see Ref.~\cite{2019PhRvD.100b3523D} for further detail). We also integrate Eq.~(\ref{nux}) to find the number density $n_\mathrm{peaks}$ of peaks.
The dotted curves in Fig.~\ref{z10}, with respect to the right-hand axis, show the redshift $z_5$ by which 5 percent of the peaks have collapsed. This number largely determines the annihilation rate in cusps; see the left-hand axis, which is connected to the right-hand axis via Eq.~(\ref{JperM}). Generally, larger particle masses and earlier decoupling cause cusps to form earlier because both effects move the free-streaming cutoff $k_\mathrm{fs}^{-1}$ to smaller scales, where there is (logarithmically) more power in density variations.

The number of cusps per dark matter mass is the ratio $n_\mathrm{cusps}/\bar\rho$ between the cusp number density and the mean dark matter density. Note that $n_\mathrm{cusps}=\fsurv f_\mathrm{coll}n_\mathrm{peaks}$, where $f_\mathrm{coll}$ is the fraction of peaks that have collapsed by the time under consideration (and recall that some peaks are taken never to collapse, owing to Eq.~\ref{ec} having no solution for them). $\fsurv$ is the fraction of peaks that are predicted to have collapsed that are associated with surviving prompt cusps; see Appendix~\ref{sec:simsurv}. The annihilation $J$ factor per mass is then $J_\mathrm{cusps}/M_\mathrm{DM} = \bar J n_\mathrm{cusps}/\bar\rho$, where $\bar J$ is the average value of Eq.~(\ref{Jcusp}) over the cusps that have collapsed. The solid curves in Fig.~\ref{z10} show $J_\mathrm{cusps}/M_\mathrm{DM}$ for a range of WIMP scenarios (assuming $\fsurv=1$). We also show (dashed curves) the fitting form in Eq.~(\ref{JperM_particle}), which is evidently accurate to about 10 percent; we separately verified its accuracy at this level for dark matter candidates with $\mathrm{MeV}<m_\chi<\mathrm{PeV}$ and $\mathrm{keV}<T_\mathrm{kd}<m_\chi$. For the solid and dashed curves, the right-hand axis of Fig.~\ref{z10} indicates the effective value of $z_5$ that makes Eq.~(\ref{JperM}) yield the correct value of $J_\mathrm{cusps}/M_\mathrm{DM}$. Generally, the effective $z_5$ grows slightly more quickly than $z_5$ because as the free-streaming cutoff $k_\mathrm{fs}^{-1}$ moves to smaller scales, the (dimensionless) power spectrum near the cutoff becomes shallower, which makes peak sizes $R$ slightly larger in relation to $k_\mathrm{fs}^{-1}$.

\section{Prompt cusp survival in simulations}\label{sec:simsurv}

As discussed in the main text, the calculation above can overestimate the number of prompt cusps and their annihilation signal because of two effects: cusps can merge onto other cusps, and peaks can fail to form cusps if they are incorporated into existing structures prior to collapse. Here we use numerical simulations to show that together these two effects reduce the number of cusps and their annihilation signal by about a factor $\fsurv\simeq 0.5$.
Due to the extreme computational requirements, we do not attempt to resolve the internal structure of prompt cusps. That would require each cusp to contain millions of simulation particles, as in the high-resolution zoom simulations of Ref.~\cite{2023MNRAS.518.3509D}. If cusps are Earth-mass, as in the 100~GeV WIMP scenario that we consider, then simulating even a solar mass of dark matter would demand as many as $10^{12}$ simulation particles, comparable to the very largest cosmological dark matter simulations carried out thus far (e.g.~\cite{2017ComAC...4....2P,2021MNRAS.506.4210I,2023MNRAS.524.2556H}). Instead, we use numerical simulations to track halos at population level, and we base our estimates on the principle that
tidal stripping always leaves central cusps intact at sufficiently small radii \cite{2021arXiv211101148A,2022MNRAS.517.1398B,2023MNRAS.tmp..828S,2022MNRAS.516..106D,2020MNRAS.491.4591E}. Such cusps can thus only be destroyed by merging into the central cusps of their (larger) host halos.

\begin{figure}
\centering
\includegraphics[width=\textwidth]{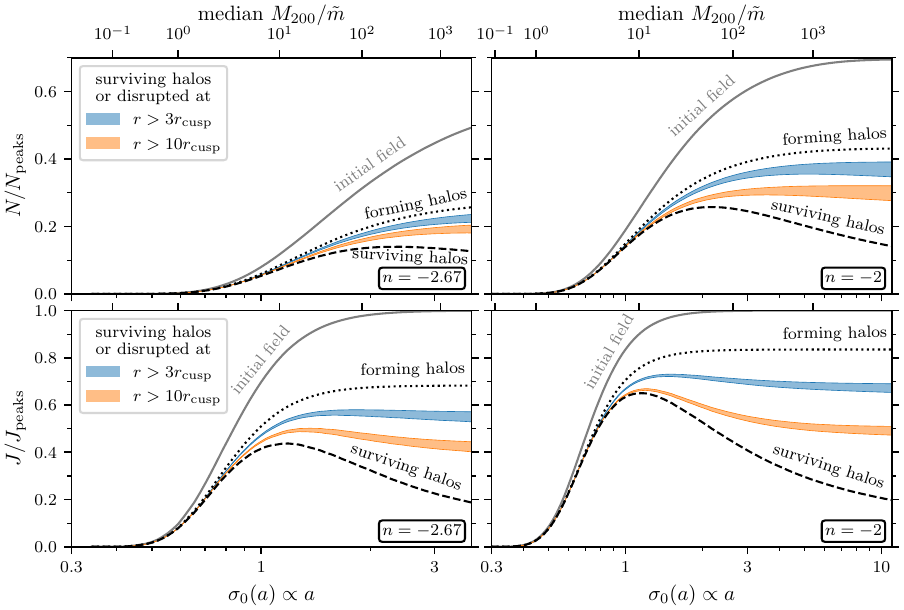}
    \caption{
    {Estimates of prompt cusp survival in numerical simulations.}
    We consider the $n=-2.67$ (left-hand panels) and $n=-2$ (right-hand panels) simulations of Ref.~\cite{2023MNRAS.518.3509D}.
    The upper panels show the number of prompt cusps, in units of the total count $N_\mathrm{peaks}$ of initial density peaks. The lower panels show the predicted annihilation rate, in units of the total predicted rate summed over all initial peaks.
    The solid curves assume that every peak in the initial conditions gives rise to a prompt cusp that survives indefinitely.
    The dotted curves are restricted to the cusps associated with halos that  can be explicitly identified as forming within the simulations. The dashed curves are further restricted to the subhalos that still survive within the simulations at each time.
    The disruption of many subhalos is, however, a numerical artifact; cuspy subhalos should survive indefinitely unless they sink to the centers of their hosts \cite{2020MNRAS.491.4591E}. The colored bands represent the prompt cusps that should survive based on this consideration; we assume a prompt cusp to survive if its subhalo disrupts before it gets closer to the center than 3 (blue band) or 10 (orange band) times the larger of the radii of the host's or subhalo's prompt cusps.
    The thickness of these bands accommodates two different assumptions about field halos that are disrupted in the simulations: either their disruption is artificial (so their prompt cusps are assumed to persist) or their formation is artificial (so their prompt cusps are neglected).
    The lower axis indicates the {\it rms} linear density contrast $\sigma_0\propto a$, which is a time parameter.
    The upper axis indicates the mass-weighted median mass of field halos, i.e. the mass of a typical halo particle's top-level host, in units of the prompt cusp mass scale $\tilde m$.
    }\label{Jsurv}
\end{figure}

We consider the $n=-2.67$ and $n=-2$ cosmological simulations of Ref.~\cite{2023MNRAS.518.3509D}, rather than the zoomed subvolumes which were followed at much higher resolution. These simulations adopted power-law initial power spectra with a Gaussian cutoff, $P(k)=k^n \exp[-(k/k_\mathrm{cut})^2]$. The cutoff wavenumber $k_\mathrm{cut}$ can be used to define the scale of the simulation. We take $\tilde m\equiv 7.3\langle k^2\rangle^{-3/2}\bar\rho$ as our mass unit, where $\bar\rho$ is the mean comoving dark matter density and $\langle k^2\rangle\equiv\int{\diff^3 \vec k}~k^2 P(k)/\int{\diff^3 \vec k}~P(k)=[(3+n)/2]k_\mathrm{cut}^2$. Note that $\tilde m$ is the mass of a typical prompt cusp \cite{2023MNRAS.518.3509D}.
The simulation particle mass and box mass are $\tilde m/440$ and $2.4\times 10^6\tilde m$, respectively, for the $n=-2.67$ simulation,
and $\tilde m/83$ and $1.3\times 10^7\tilde m$, respectively, for the $n=-2$ simulation. Both simulations begin with $1024^3$-particle grids, perturbed using second-order Lagrangian perturbation theory, at a time when the rms density contrast $\sigma_0\equiv [\int\frac{\diff^3 \vec k}{(2\pi)^3} P(k)]^{1/2}=0.03$. They are evolved under gravity using the \textsc{Gadget-4} simulation code \cite{2021MNRAS.506.2871S} with a comoving force-softening length equal to 0.03 times the grid spacing. Simulation snapshots are separated by a factor of about 1.035 in $a$.
At the final times, half of all halo mass resides in individual halos with $M_{200}/\tilde m> 1500$ and $8500$ in the $n=-2.67$ and $-2$ simulations, respectively (see the upper axes of Fig.~\ref{Jsurv}), where $M_{200}$ is the mass within the radius $R_{200}$ that encloses 200 times the cosmological mean density.
The initial conditions of the $n=-2.67$ and $-2$ simulations contain, respectively, $N_{\rm peaks} = 9.1\times 10^5$ and $1.45\times 10^6$ peaks in total.

The solid curves in Fig.~\ref{Jsurv} indicate the fraction of all these initial peaks which are predicted to have collapsed by each expansion factor, together with the fraction of the final total annihilation luminosity predicted to come from their prompt cusps.
Although these simulations cannot resolve the inner structures of halos, they can track halo populations.
Specifically, we consider halos (both field halos and subhalos) identified by \textsc{Gadget-4}'s \textsc{subfind-hbt} algorithm.
For each local maximum in the initial density field, we identify the 7 nearest particles and find the first time later than the peak's predicted collapse time $\ac$ at which at least 4 of those particles reside in the same \textsc{subfind-hbt} halo. If it does not already contain a prompt cusp from a different, previously collapsed peak, that halo is then deemed to be the host of the prompt cusp that forms from the initial density peak under consideration. If, however, this halo does contain a previously formed cusp, then the peak under consideration is considered not to form its own halo but rather to be incorporated pre-collapse into the other object.
If we find that two or more peaks are first incorporated into the same, previously ``cusp-free'' halo at the same snapshot, then we consider the one with the earliest predicted collapse time to produce the ``true'' cusp and we reject the others.
The dotted curves in Fig.~\ref{Jsurv} are for those cusps whose halos actually form in the simulations according to this definition. A significant fraction of the lower (and thus later-collapsing) peaks are predicted not to form their own halos or cusps.\footnote{
In principle, it would also be insightful to show in Fig.~\ref{Jsurv} the total number of halos identified in each simulation. However, that count is inflated by the artificial fragmentation of filamentary structures into bound clumps (e.g.~\cite{2007MNRAS.380...93W,2013MNRAS.434.3337A,2014MNRAS.439..300L}), such that a significant fraction of simulation-identified halos are numerical artifacts. Since nonlinear structure is expected to originate with the collapse of initial density peaks, we assume that only the simulation-identified halos that arise from initial density peaks are real.
}

Each halo in one snapshot is linked to its primary descendant in the next snapshot by \textsc{Gadget-4}'s merger tree algorithm, so with these links, we have the full history of the halo associated with each initial peak. In cases where a halo has no descendent in the next snapshot, we follow the halo's most-bound particle until it becomes a member of another \textsc{subfind-hbt} halo, and that halo (if it exists) is then considered to be the descendent. As structure builds up, halos merge onto other halos and most become subhalos of larger objects.
If the central cusps of  \textsc{subfind-hbt}-identified halos and subhalos are considered to be the only surviving prompt cusps,  then the total number of cusps and their total annihilation luminosity are dramatically reduced at later times, as indicated by the dashed curves in Fig.~\ref{Jsurv}.
However, much of the reduction relative to the dotted curves is a result of the artificial disruption of simulated subhalos due to the limited spatial and mass resolution of the simulations. As noted above, a real subhalo's central cusp can only be destroyed if it sinks into the center of its host. When subhalos disrupt in the simulations, we therefore record their distance from host center; specifically, we measure the distance to  host center at the last snapshot when the subhalo is identified by \textsc{subfind-hbt}. If this distance exceeds some predetermined value $r_{\rm max}$, we assume the disruption to be artificial and all the prompt cusps associated with the subhalos  are assumed to survive to later times. If disruption occurs at $r<r_{\rm max}$, we assume that the central cusp of the subhalo is destroyed but that all other prompt cusps associated with it survive.
The colored bands in Fig.~\ref{Jsurv} show the number and predicted annihilation rate of cusps whose subhalos either survive in the simulation or are disrupted at excessive distance from their hosts. We consider two distance thresholds, $r>3\Rc$ (blue) and $r>10\Rc$ (orange), where $\Rc$ is the larger of the radii of the two prompt cusps concerned. In principle, we would expect the relevant threshold to be $\sim \Rc$, but the finite time resolution between simulation snapshots means that the moment of closest approach can be missed. Additionally, it seems sensible to be conservative, since a cusp may sink further towards host center, and so eventually be absorbed, after its halo is artificially disrupted in the simulations.

There is one further complication, which is why we plot the populations described above as finite bands in Fig.~\ref{Jsurv}. Some initial peaks are associated with halos that have no descendent beyond some time, even after tracking the future halo membership of the most-bound halo particle as described above. This phenomenon means that a halo was disrupted in the field in the simulation (and not as a subhalo, because then the subhalo's host would be the descendent). There are two ways to interpret such an event. There is no obvious physical mechanism to disrupt a field halo, so the disruption could be considered artificial. In that case, the prompt cusps associated with these halos should be assumed to survive, yielding the upper edges of the colored bands in Fig.~\ref{Jsurv}. However, since it is also unclear how a field halo would be artificially disrupted, another interpretation could be that these objects were misidentified as bound halos by the group-finding algorithms. In this case, their associated prompt cusps would also be artificial. Deleting them yields the lower edges of the colored bands in Fig.~\ref{Jsurv}. The narrowness of the bands shows that this ambiguity is not a major concern.

The main conclusion to be drawn from the lower panels of Fig.~\ref{Jsurv} is that at late times, the predicted annihilation luminosity from surviving prompt cusps is about half that inferred from the full set of initial peaks. This conclusion varies moderately with the cusp disruption threshold $r_{\rm max}$ (different colors). The number of surviving cusps (upper panels) is also about half the number of initial peaks predicted to collapse. It is important that $J/J_\mathrm{peaks}$ approaches a steady value at later times, because the requirement that the scale of collapsed structures not approach the size of the simulated region allows only a comparatively short time interval to be followed. The late-time convergence of $J/J_\mathrm{peaks}$ suggests that this does not adversely affect our conclusions. This late-time value is lower by about 0.1 in the $n=-2.67$ simulation than in the $n=-2$ simulation. The power spectrum near the cutoff scale for a 100~GeV WIMP model has $n\sim -2.8$, which is more difficult to simulate due to the rapid growth of collapsed structure. However, given the small difference in final $J/J_\mathrm{peaks}$  between $n=-2$ and $n=-2.67$, there is little reason to expect a large change as $n\to -2.8$.
By the end of the $n=-2.67$ simulation, half of the surviving prompt cusps lie in halos which individually contain at least 4 to 10 cusps, depending on $r_{\rm max}$ and whether cusps from disrupted field halos are included.
Half of their annihilation luminosity comes from halos containing at least 600 to 800 cusps.
For the $n=-2$ simulation, the corresponding numbers are 30 to 80 for cusp number and 500 to 900 for annihilation luminosity. Thus, despite the restricted dynamic range of our simulations, halos containing many prompt cusps already dominate the annihilation luminosity at the final time.

We can also discuss the relative contributions of the two main mechanisms that reduce the cusp count below the predictions from the initial conditions (the solid lines in Fig.~\ref{Jsurv}). 
The difference between the solid curves and the dotted curves is due to initial peaks which are deemed never to form cusps, while the difference between the dotted curves and the colored bands is due to cusps which do form but are later destroyed by merging. Evidently, cusp nonformation is the dominant effect suppressing the cusp count (upper panels), but both effects impact the annihilation rate (lower panels) to comparable degrees. The impact of cusp nonformation tends to be larger, and the impact of cusp disruption smaller, for the $n=-2.67$ case, compared to the $n=-2$ case.

\section{Tidal stripping of prompt cusps}\label{sec:tidal}

Near the center of the Galactic halo and other dark matter halos, prompt cusps can be gradually stripped by tidal forces, an effect that reduces their annihilation signal. Previous numerical studies have explored the impact of tidal evolution on subhalo annihilation rates \cite{2019PhRvD.100f3505D,2023MNRAS.518...93A}, but they assumed NFW density profiles, so their results cannot be directly applied to prompt cusps. For example, a prompt cusp's annihilation rate is only logarithmically sensitive to its outer radius, and so it is only weakly affected by loss of outer material. Also, steeper cusps are more resistant to tidal effects than those of NFW haloes \cite{2010MNRAS.406.1290P}.

However, several theoretical models of tidal evolution have been developed \cite{2021arXiv211101148A,2022MNRAS.517.1398B,2022MNRAS.516..106D,2023MNRAS.tmp..828S}. All of these models predict that a central cusp should remain intact at sufficiently small radii, even as a subhalo's outer regions expand and are stripped.
We use the \textsc{adiabatic-tides} model \cite{2023MNRAS.tmp..828S} to evaluate the asymptotic tidal remnant of an idealized $\rho\propto r^{-1.5}$ power-law density profile (which has no truncation or core radius). Since the idealized profile is scale-free, the result can be straightforwardly rescaled to apply to any prompt cusp in any region of any halo. To appropriately account for the core radius $\rc$ and the cusp's outer radius $\Rc$, we next scale the tidally stripped idealized profile by the ratio between the cusp's initial profile and the initial idealized power law.
This procedure is not exact, but we have verified that alternative procedures (like truncating at the cusp mass $\Mc$ instead of the cusp radius) do not change the resulting annihilation rate significantly.

The degree of stripping in the \textsc{adiabatic-tides} model depends on the subhalo's pericenter radius.
Within an isothermal sphere, which has the density profile $\rho\propto r^{-2}$ and is typically a good approximation for a large range of radii inside Galactic systems, the average pericenter radius of particles at radius $r$ is about $r_\mathrm{peri}\simeq 0.55r$.\footnote{
For a particle at radius $r$ and velocity $\vec v$, energy conservation implies that $r_\mathrm{peri}$ satisfies $\Phi(r_\mathrm{peri})+L^2/(2r_\mathrm{peri}^2)=\Phi(r)+v^2/2$, where $\Phi$ is the potential and $L=v_\mathrm{t}r$ is the angular momentum, which depends on the tangential component $v_\mathrm{t}$ of the velocity $\vec v$.
For the isothermal sphere potential $\Phi(r)=2\sigma^2\log(r)$ (where $\sigma$ is the velocity dispersion per dimension), this equation becomes $2\sigma^2\log(r_\mathrm{peri}/r)+(r_\mathrm{peri}/r)^{-2} v_\mathrm{t}^2/2=\vec v^2/2$, which can be solved numerically to yield $r_\mathrm{peri}/r$ as a function of $\vec v$.
We can then integrate $r_\mathrm{peri}/r$ over the velocity distribution $f(\vec v)=(2\pi\sigma)^{-3/2}\e^{-\vec v^2/(2\sigma^2)}$; the numerical result is that $\langle r_\mathrm{peri}/r\rangle\simeq 0.55$.
} The average value of $r_\mathrm{peri}/r$ is higher for the shallower profiles relevant to halos' deep interiors. We therefore conservatively approximate a prompt cusp's pericenter radius as half of its current radius, so prompt cusps at the radius $r$ are assumed to have pericenters $r/2$.
The solid curve in Fig.~\ref{boost} shows annihilation boost factors when tidal stripping is taken into account.

\section{Prompt cusp signal associated with the Galactic Center excess}

We adopt the same model of the Galactic dark matter halo as Ref.~\cite{2023MNRAS.523.1067S}. The halo is taken to initially have an NFW density profile with concentration $c\equiv R_{200c}/r_s=8.7$ and mass $M_{200c}=10^{12}$~M$_\odot$, but it is adiabatically contracted due to the baryonic components according to the prescription of Ref.~\cite{2020MNRAS.494.4291C}. These baryonic components include the axisymmetric stellar and gas disks and a stellar bulge as in Ref.~\cite{2019MNRAS.487.4409K}, which uses parameters consistent with observational constraints \cite{2016ARA&A..54..529B,2017MNRAS.465...76M}.
In order to attribute Ref.~\cite{2021PhRvD.103f3029D}'s measurement of the Galactic Center excess to dark matter annihilation in this smooth Galactic halo, the differential photon power emitted by a mass $M_\mathrm{DM}$ of dark matter with density $\rho$ must be
\begin{equation}
    E^2\frac{\diff N}{\diff E} = \mathcal{N}_\mathrm{GCE}(E)
    \frac{M_\mathrm{DM}}{\mathrm{M}_\odot}
    \frac{\rho}{\mathrm{M_\odot}\,\mathrm{Mpc}^{-3}}.
\end{equation}
Here,
\begin{equation}
    \mathcal{N}_\mathrm{GCE}(E) =
    8.6\times 10^{14}\ \mathrm{MeV}\,\mathrm{s}^{-1}
    \left(\frac{E}{\mathrm{GeV}}\right)^{0.27-0.27\log\left(\frac{E}{\mathrm{GeV}}\right)}
\end{equation}
is a rescaled form of the log-parabola fit by Ref.~\cite{2021PhRvD.103f3029D} to the energy spectrum of the $\gamma$-ray excess. For prompt cusps, we replace $\rho$ with the effective density $J_\mathrm{cusps}/M_\mathrm{DM}$.

For a line-of-sight angle $\Omega$, the energy flux from a low-redshift galaxy or cluster halo, such as our own Galactic halo, is
\begin{equation}\label{flux}
    E\frac{\diff^2\Phi_E}{\diff\Omega \diff E}=
    \frac{1}{4\pi}\mathcal{N}_\mathrm{GCE}(E)
    \int_0^\infty\diff\ell
    \left[\frac{J_\mathrm{cusps}}{M_\mathrm{DM}}\rho(\ell,\Omega)+\rho(\ell,\Omega)^2\right],
\end{equation}
where $\rho(\ell,\Omega)$ is the density of the smooth halo at the position $\ell$ along the line of sight. The sum accounts for both the smooth halo and the prompt cusps.
For our own Galactic halo, $J_\mathrm{cusps}/M_\mathrm{DM}$ depends significantly on position due to influence of tidal forces and stellar encounters, and we use the \textsc{cusp-encounters} code to evaluate these effects \cite{2023MNRAS.523.1067S}.
Meanwhile, from unresolved extragalactic structures, the average contribution to the energy flux is cosmologically redshifted and is given by
\begin{equation}\label{egflux}
    E\frac{\diff^2\Phi_E}{\diff\Omega \diff E}=E^{2}
    \frac{c}{4\pi}\rhoc
    \frac{J_\mathrm{cusps}}{M_\mathrm{DM}}
    \int_0^\infty\frac{\diff z}{H(z)}
    \left[\frac{\mathcal{N}_\mathrm{GCE}(E^\prime)}{E^{\prime 2}}\right]_{E^\prime = E(1+z)}.
\end{equation}
Here we neglect scattering with extragalactic background light, which is minor at these energies.

\begin{figure}[t]%
\centering
\includegraphics[width=0.7\textwidth]{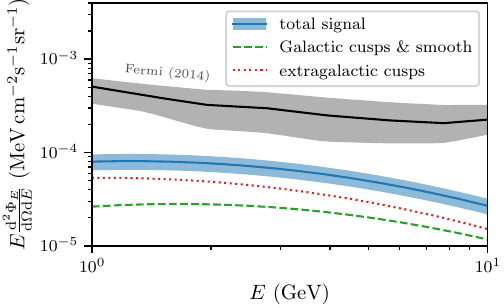}
\caption{
{Energy spectra associated with the contribution of dark matter annihilation to the isotropic $\gamma$-ray background if the Galactic Center $\gamma$-ray excess is due to annihilation.}
We show the measured differential energy flux (black curve with uncertainty band) along with the predicted contribution of the Galactic halo (dashed curve), that of extragalactic dark matter (dotted curve), and the sum of the two contributions (blue solid curve). The $\fsurv=0.5\pm 0.1$ uncertainty band accompanies the last curve.}\label{spectrum}
\end{figure}

Figure~\ref{spectrum} shows the differential energy flux predicted by Eq.~(\ref{flux}) for the Galactic halo (dashed curve) and Eq.~(\ref{egflux}) for the averaged extragalactic field (dotted curve). The Fermi Collaboration's measurement of the isotropic $\gamma$-ray background \cite{2015ApJ...799...86A} is also plotted for comparison.
For the Galactic contribution, the signal is averaged over Galactic latitudes $|b|>20$~degrees, since that is the latitude cut used for the measurement.
For Figs. \ref{galactic} and~\ref{clusters}, we integrate over the 1-10~GeV energy range and consider how the integrated flux depends on angle.

\bibliography{refs.bib}

\end{document}